# Correcting Presbyopia with Autofocusing Liquid-Lens Eyeglasses

Mohit U. Karkhanis[†], Chayanjit Ghosh[†], Aishwaryadev Banerjee[†], Nazmul Hasan[§], Rugved Likhite[§], Tridib Ghosh[*], Hanseup Kim[†] and Carlos H. Mastrangelo[†]

Mohit U. Karkhanis[†], Chayanjit Ghosh[†], Aishwaryadev Banerjee[†], Nazmul Hasan[§], Rugved Likhite[§], Tridib Ghosh[*], Hanseup Kim[†] and Carlos H. Mastrangelo[†]

*Abstract*— Presbyopia, an age-related ocular disorder, is characterized by the loss in the accommodative abilities of the human ocular system and afflicts more than 1.8 billion people world-wide. Conventional methods of correcting presbyopia fragment the field of vision, inherently resulting in significant vision impairment. We demonstrate the development, assembly and evaluation of autofocusing eyeglasses for restoration of accommodation without vision field loss. The adaptive optics eyeglasses consist of two variable-focus piezoelectric liquid lenses, a time-of-flight range sensor and low-power, dual microprocessor control electronics housed within an ergonomic frame. Patient-specific accommodation deficiency models were utilized to demonstrate a high-fidelity accommodative correction. Each accommodation correction calculation was performed in ~67 ms requiring 4.86 mJ of energy. The optical resolution of the system was 10.5 cycles/degree, featuring a restorative accommodative range of 4.3 D. This system can run for up to 19 hours between charge cycles and weighs ~132 g, allowing comfortable restoration of accommodative function.

## I. Introduction

PRESBYOPIA is an age-related condition, which stems from the gradual loss in the eye's ability to change its optical power [1]. It manifests as an inability to focus on objects, often leading to symptoms like visual discomfort, eye strain and headache, after around 45 years of age [2]. The amount of accommodation loss as a function of age is characterized by Duane's curve [3], shown in Fig. 1. This curve shows the accommodative range (or accommodative amplitude AA) of the eye crystalline lens versus age. A young person's crystalline lens has about 12 D of accommodative range, but that range is reduced to about 1 D by age 45 (physically, this translates into being able to focus objects located between 0 m and 1/AA m). A young individual can thus see objects clearly between infinity up to 8 cm in front of them. An older, 60-year-old presbyope can only see objects between infinity to 1 m or worse as loss of accommodation often is accompanied by a fixed power offset.

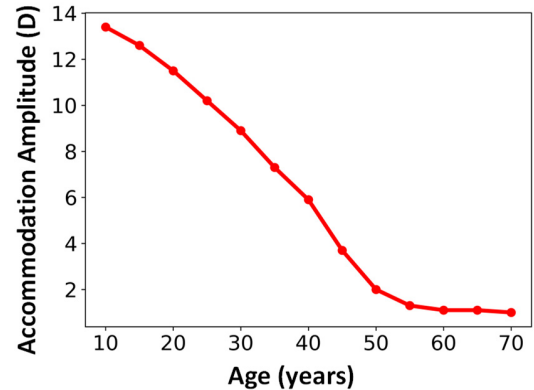

Fig. 1. Representation of Duane's curve of accommodation loss versus age [3]. The amplitude of accommodation progressively reduces to <1.0 D as a person enters their fifth decade. Note that the curve flattens out but does not become zero indicating that some degree of remnant accommodation is present.

The precise mechanism underpinning the development and progression of presbyopia is still debated, but the principal reasons behind this condition are the loss of power in the ciliary muscles and the morphological changes in the crystalline lens of the eye [3]–[22]. In 2018, globally, more than 1.8 billion people were afflicted with this condition [23]. With increasing life expectancies, the number of people affected by this debilitating condition is expected to keep on rising [24]. Currently presbyopia can be partially corrected using several optical devices and methods listed below, none of which can restore normal vision.

*Multifocal Corrective Eyeglasses:* Traditional methods used to correct vision loss due to presbyopia include the use bifocal, trifocal, multifocal and progressive eyeglasses [25]. These devices feature lenses which divide the visual field into zones of different optical powers. The lens zoning enables a user to focus on near or far objects by adjusting their viewing angle. However, these mature technologies have serious drawbacks. Bifocal and trifocal eyeglasses provide a very fragmented field of view per zone which causes "image jumps" when the user adjusts their viewing angle [26]. Design of highly customized progressive lenses is now possible using free-form surfacing and sophisticated computer-aided modelling [27]–[29]. However, such lenses are expensive, have a steep learning

Manuscript submitted 1/21/2021.
This work was supported by the US National Institutes of Health NIBIB 1U01EB023048-01 cooperative agreement. (Corresponding author: Carlos H. Mastrangelo. email: carlos.mastrangelo@utah.edu)

† Department of Electrical and Computer Engineering, University of Utah, Salt Lake City, UT, USA.
§ Intel Corporation, Portland, OR, USA.
* NewEyes Inc., Salt Lake City, UT, USA.

curve and come with a penalty of unwanted astigmatism which distorts the user's peripheral visual field [30]–[32]. Presbyopes who use multifocal eyeglasses are also more likely to experience accidents and mobility issues due to dioptric blurs in their visual field [33]–[39].

*Monovision and Multifocal Contact Lenses:* The idea of correcting presbyopia using contact lenses has been around since the last five decades [40] but it has enjoyed limited success [26], [41], [42]. Presbyopic contact lens technologies have been extensively reviewed in literature [43]–[48] and can be divided into two main categories- monovision and multifocal contact lens correction. In monovision correction, the dominant eye is usually corrected for distance vision, while the other is corrected for near vision [49]–[54]. Many clinical trials have shown that after an adaptation period to monovision correction, presbyopes experienced an improved range of clear focus, at the expense of reduced contrast sensitivity and stereopsis [55]–[64]. There are also hard limitations on the optimum optical power of the near addition lens in monovision correction, as higher optical powers negatively impact stereopsis [49]–[51], [65]–[67]. Studies report that presbyopes successfully adapt to multifocal contact lenses despite the presence of optical aberrations [68]–[74]. Despite such successes, contact lens correction of presbyopia remains unpopular due to discomfort and inconvenience, poor visual experience, and cost [75], [76].

*Intra-Ocular Lens Replacement:* Many surgical approaches have been suggested for the correction of presbyopia, which involve replacing the natural crystalline lens of the eye with an artificial intra-ocular lens (IOL). These IOLs generally follow the same principles of refraction as the monovision and multifocal contact lenses mentioned above [47], [77]. However, this correction strategy requires detailed knowledge of the patient's ocular system, involves very precise lens calibrations before surgery, and a very delicate surgical procedure to implant the lens in the eye. Successes of this correction strategy are debatable as patients have reported haloes and glare in their vision and low contrast sensitivity [78].

*Accommodating Lenses:* A major problem with all the above methods is that they only provide a partial and unnatural tiled or warped correction of vision. To restore normal vision, the accommodation must be corrected actively, which means it must be dependent on the observed object distance, without any losses in the visual field or viewing angle adjustments. Restoring active accommodation in such patients, requires a variable optical power change of at least 3.0 D [26]. The earliest practical attempts at developing variable-focus eyeglasses were spearheaded by B. F. Edwards with his "polyfocal spectacles" in the 1950s [79]. These eyeglasses featured variable-focus, fluid-filled lenses [80] made from deformable membranes, a reservoir to store the optical fluid and a mechanism to "actuate" or control the amount of fluid in the variable-focus lens, thereby changing its optical power. Improvements in the actuation mechanism of such variable-focus lenses have been suggested by multiple inventors and researchers [81]–[84]. J. D. Silver's pioneering work to provide low-cost, variable-focus eyeglasses [85] was taken further by Adlens Ltd. with their adaptive eyewear products [86]. In addition to liquid-filled lenses, such variable-focus lenses can be implemented utilizing other technologies such as the sliding Alvarez lenses [87], electro-wetting lenses [88] and liquid crystal (LC) lenses [89]–[92], each holding its own set of advantages and pitfalls [93], [94]. The fabrication of slim, lightweight variable-focus lenses with 30-50 mm apertures necessary for eyeglasses, however, still remains a challenge.

*Smart Autofocusing Wearables:* Recent advances in fluid-filled lenses [95]–[98] have enabled successful integration of large aperture variable-focus lenses into eyeglasses [98]–[101] integrated with object range sensors which measure the distance between the eyeglasses and the object a patient is trying to focus on. The range measured is utilized to automatically configure the power of the variable-focus lenses that corrects the accommodative deficiency [98], [99], as shown in the schematic of Fig. 2. Previously described unique designs [100],

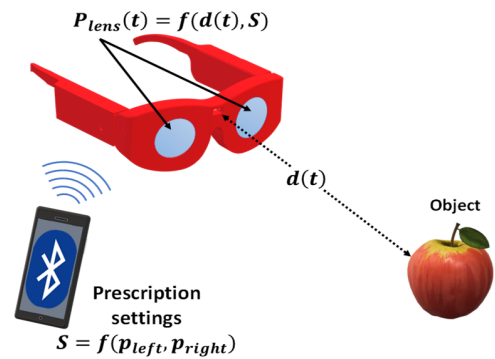

Fig.2. Schematic of a smart eyeglasses system. The adaptive lenses continuously adjust the lens power to bring the object in focus for the observer along the line of sight. The lenses must provide the accommodation deficiency, a function of the object distance.

[101] feature binocular eye tracking systems to analyze the gaze of the wearer and vary the optical power of the variable-focus lens. Such systems require investments in expensive computational devices and smartphones. However, authors in these articles show through clinical trials, that variable-focus eyeglasses outperform the traditional methods of correcting presbyopia.

A successful implementation of variable-focus eyeglasses requires the autofocusing system to reliably tune correct optical powers with minimal delay, while at the same time, crucial factors such as the system's electrical power consumption, affordability and aesthetics cannot be ignored. This paper attempts to provide a solution to the drawbacks of the existing methods of alleviating presbyopia. We utilize the adaptive optics technology developed by Hasan et.al. [95] to implement autofocusing eyeglasses. We demonstrate the performance of this system in terms of its ability to provide automatic, adequate optical correction to a presbyopic eye at 5 different distances ranging from 1 m to 30 cm, as presbyopes usually find it difficult to focus on objects placed within 1m. We discuss the electrical performance of this system and its ability to operate without an external power source. We also demonstrate the optical performance of this system using a 1951 USAF optical resolution test chart and we compare the modulation transfer function (MTF) of our variable-focus lens to that of an average human eye.

## II. MATERIALS AND METHODS

### A. System Assembly

The housing for the eyeglasses was designed in Blender 2.79b, an open source, 3D creation platform. This design was 3D printed at a 3D printing service provider (Xometry) using a nylon-based polymer and selective laser sintering (SLS) process. Electronic PCBs were designed in the electronic design automation (EDA) software, Eagle 9.6.2. These designs were manufactured on low profile, two-layered PCBs. Surface mount technology (SMT) components were used while designing the PCBs to keep their height as small as possible. Flat pack connectors (FPC) and cables were used to facilitate communications between different PCBs. Software necessary for the proper functioning of the eyeglasses, including patient-specific algorithms were programmed into the microcontrollers with Arduino integrated development environment (IDE). Finally, the lenses and the PCBs were fixed in their respective housing with the means of micro screws.

### B. Optical Measurements

The performance of the optical components and subsystems of the set were measured using the following methods.

*Accommodation vs Object Distance:* The performance of the variable-focus lenses, being the most critical subsystem within the autofocusing eyeglasses, were evaluated for their optical response with respect to different actuator voltages. The optical power of each lens was measured using a Shack-Hartman wavefront sensor (Thorlabs WFS150-7AR) while varying the bimorph actuator voltage from -200V to +200V. This process was repeated 5 times.

To evaluate the system response (i.e., lens optical response) in terms of the accommodation stimulus (reciprocal of the object distance), an object was placed at 5 different distances (1 m, 70 cm, 50 cm, 40 cm, 30 cm) in front of the autofocusing system, as shown in Fig. 3. The ToF distance sensor and control electronics were attached to the same lens-holder which held the variable-focus lens. This allowed the distance sensor to gather real time distance data of an object placed directly in its line of sight and communicate this distance to the control electronics which set the corrective optical powers on the variable-focus lens. The relationship between the object distance and the lens corrective power was determined using patient-specific empirical accommodation deficiency models [102] programmed into the control electronics. Patient data necessary to test these eyeglasses were obtained from a recent, unpublished clinical trial study undertaken by authors of this article [102], [103]

We used a 633 nm light source and the Shack-Hartman wavefront sensor to measure the various optical powers generated by this system at various object distances, as shown in the schematic of Fig. 3. The distance sensor and the control electronics were attached to the variable-focus lens holder. The object was moved towards or away from the distance sensor while the autofocusing system automatically changed the optical power of the lens in response to the object distance. The minimum and maximum object distances were 30 cm and 1 m, respectively.

*Optical Resolution:* To evaluate the approximate optical resolution of the variable-focus lens undergoing automatic focus correction, we repeated the experiment described above while utilizing a 1951 USAF optical resolution test chart. This test chart was placed at the 5 different distances from the eyeglasses, mentioned above, and a DSLR camera (Canon EOS 1200D) was used to obtain raw images of the resolution test chart through the tuned variable-focus lenses. Test chart illumination was kept constant using a diffused, LED studio lighting system. The correlated color temperature of the lighting system was kept constant at 5000 K. The DSLR camera's ISO, aperture and shutter speed were fixed at 1600, *f*/22 and 0.5 s, respectively. Fig. 4 shows a schematic of the setup used for this experiment.

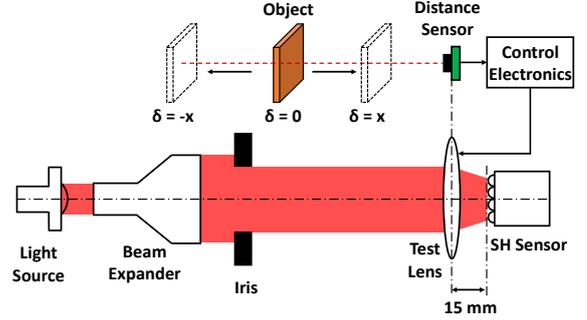

Fig. 3. Schematic of the experimental setup used to measure the power –distance relation of the variable-focus lens.

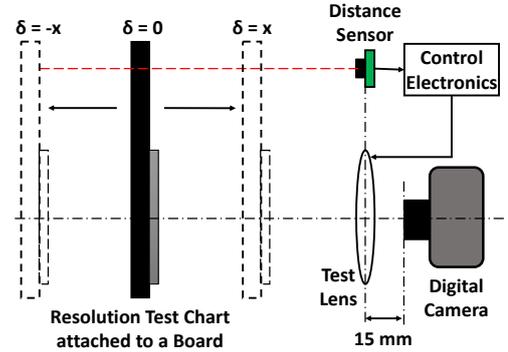

Fig. 4. Schematic of the experimental setup to measure the optical resolution of the variable-focus lens. The 1951 USAF resolution test chart was mounted on a board.

The distance sensor and control electronics were attached to the variable-focus lens holder. The board was moved towards and away from the distance sensor, with minimum and maximum board-distance sensor distances being 30 cm and 1 m, respectively. The autofocusing system automatically tuned the variable-focus lens in response to the board distance.

*Modulation Transfer Function (MTF):* The MTF of an optical system is its intensity response as a function of spatial frequency [104] and a measure of the system optical quality. We estimated the 1-D MTF along the X-axis and Y-axis of the variable-focus lens used in the autofocusing eyeglasses directly from the aberration Zernike polynomial coefficients up to vertical quadrafoil (OSA/ANSI index Z14) [105], [106], which were measured with the help of the Shack-Hartman sensor. These estimated MTFs were then compared to those of an average human eye with different pupil diameters.

## III. RESULTS AND DISCUSSION

### A. System Design

It is important to note that as a person ages, their accommodative ability is reduced, but not completely lost [107], [108]. Hence, remnant accommodation plays a crucial role when a presbyope is trying to achieve clear focus. In order to restore normal vision, the variable-focus lens must reproduce the *accommodation deficiency* (AD) curve [102] as shown in Fig. 5, which is the difference between the expected ideal response of the accommodative system (clear focus) and the degraded *accommodation response* (AR) of the presbyopic eye.

The AD must be provided by the accommodative optics in order to restore normal vision. Therefore, in practical terms, the patient's accommodation deficiency data should be programmed into an individual's autofocusing eyeglasses. Such data are a function of the *accommodation stimulus* (AS), which is the reciprocal of the distance between the eyes and the object plane where clear focus is expected.

In order to compensate for the AD, the autofocusing eyeglasses thus require a distance or a depth sensor that measures the range between the observer and the object being imaged. The autofocusing eyeglasses discussed in this paper thus consists of four key parts: (1) variable-focus lenses, (2) distance/ depth sensor, (3) control electronics, and (4) patient-specific AD models. Each of these subsystems is discussed below.

*1) Patient-Specific AD Models*

The AD curves vary quite a bit between presbyopes and, while fundamentally one can perform a full measurement of the AD and program it into the autofocusing eyeglasses, the cost of such a test can be prohibitive and a barrier to the use of the autofocusing system. Therefore, we chose to use model functions for the AD that require just a few parameters (and measurements), to describe the AD more economically. Multiple accommodation response models, which express the accommodation response of an ocular system in terms of the object distance, have been suggested so far [102], [109].

We use a newly developed model described in [102], which relates the accommodation response to the object distance using a sigmoid function:

$$AR(AS, a, b, k) = \frac{a}{1 + e^{(-k(AS-b))}} \quad (1)$$

where $AR$ is the accommodation response, $AS$ is the accommodation stimulus (reciprocal of the object distance), $a$ determines the accommodative amplitude of the individual i.e., the useful linear range of the experimental accommodation response curve, $k$ determines range of stimuli for which the $AR$ curve exhibits a linear response and parameter $b$ corresponds to the midpoint of the non-flat accommodation stimulus range. It should be noted that the parameters $a$, $b$ and $k$ are patient-specific and vary from eye to eye.

This model allows us to predict how a patient's accommodation system behaves as an object is moved within their visual field. The solid curve of Fig. 5 shows a fit of the model to an experimental AD data. Accommodation can then be restored in patients by simply adding the deficiency in their accommodation system's behavior with the help of the distance data from the distance sensor and properly tuning lenses thereafter.

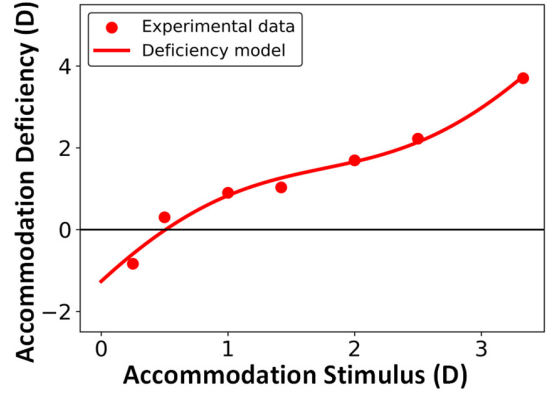

Fig. 5. Example presbyope accommodation deficiency versus accommodative stimulus curve. Accommodation stimulus is the reciprocal of the object distance from the eyes, while the accommodation deficiency is the defocus error present in the human ocular system. The solid line represents a typical AD model, developed in a recent study [102].

*2) Variable-focus lenses*

We utilize two squeezable-type liquid-filled lenses which do not require external fluid reservoirs. This significantly reduces the weight and profile of the variable-focus lens subsystem. Each lens consists of a cylindrical liquid-filled cavity, bound by two flexible polydimethylsiloxane (PDMS) membranes. Additional details of this lens, including its structure and electro-optical performance, have been described in [95].

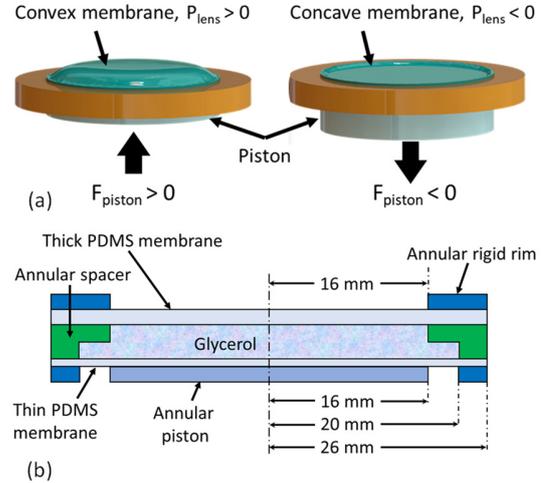

Fig. 6. (a) Working principle of liquid-filled variable-focus lens. (b) Simplified schematic of the variable-focus lens without the piezoelectric bimorph actuators [95].

Fig. 6 (a) and (b) show the working principle and a simplified cross-section of this variable-focus lens without the actuators, respectively. Three curved piezoelectric bimorphs, placed along the periphery of the lens, form the actuators. One end of these bimorphs is fixed to the lens rim, while the other is connected to a hollow piston with three extended arms. The piston is attached to one of the deformable membranes. When voltage is applied to the actuators, they move the piston into or out of the lens, depending on the applied voltage amplitude and polarity. This makes a concave or a convex profile on the other membrane, changing the optical power of the lens. Fig. 7 (a) shows a photograph of a 30 mm aperture diameter tunable liquid lens. The lens weighs less than 15 g, making them suitable for eyeglasses applications, much lighter than other piezoelectric liquid lenses (eg. Optotune EL-16-40-TC [110]). Fig. 7 (b) shows the optical response of the lens as a function of the applied voltage across the bimorph actuator. This lens features a repetitive linear optical response to the applied voltage, which greatly simplifies the electronics system required to control the lens. The response time of this lens is ~40 ms with a variable-focus range of ~4.3 D.

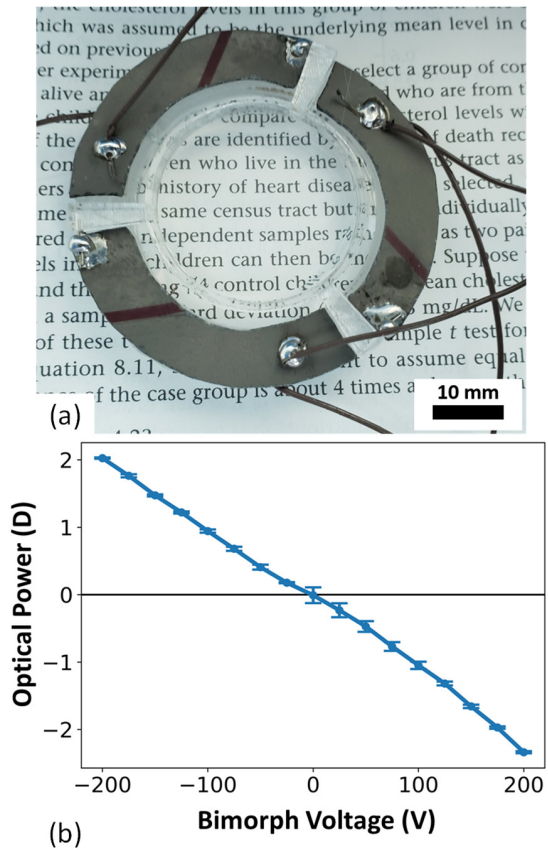

Fig. 7. (a) Photograph of the variable-focus lens [95]. (b) Optical response of the variable-focus lens as a function of bimorph actuator voltage. Bars represent the standard deviation. The optical power of the lens is a linear function of the applied voltage.

*3) Object Distance Sensor and Range Cutoff*

The role of a distance sensor in this system is crucial and requires precise measurement of the distance between the eyeglasses and the object to be focused. Distance sensors are characterized by their resolution, refresh rates (time taken per measurement) and the maximum range of reliable operation. IR (infrared) time-of-flight (ToF) distance sensors (eg. RF Digital Simblee RFD77402) provide very fine resolution measurements, with rapid refresh rates and a reasonable working range of up to 2 m, corresponding to an accommodation stimulus of 0.5 D. The autofocusing system regards objects located at distances > 2 m as being reasonably well focused at the 0.5 D cut off. This is a reasonable first order approximation as the average RMS aberration of the human eye [111]–[115] translates to an equivalent defocus error of 0.05 D-0.49 D; thus lens adjustments smaller than about 0.5 D are marginally noticed.

*4) High-Voltage Circuits and Feedback Control Electronics*

Fig. 8 shows a block diagram of the feedback control electronics. As evident from Fig. 7 (b), the variable-focus lenses used within this system require a variable voltage between +200 V and -200 V. Efficient high-voltage DC-DC converters are necessary to convert the on-board battery voltage of 3.3 V to +200 V. We use a high voltage, ultra-miniature DC-DC piezo transformer for this operation. The bimorph actuators of each lens are connected to the high voltage converter in a H-bridge configuration with the help of four high-voltage optoelectronic switches (Ixys LAA100P). These optoelectronic switches help in setting the proper polarity of the voltage across the lenses. The high-voltage converter is operated within a feedback loop, such that it does not need to be switched on all the time, which increases the electrical power efficiency of this system.

We use two microcontrollers in a master-slave configuration to control the autofocusing eyeglasses. The master microcontroller continuously fetches distance data from the ToF distance sensor, through an I$^2$C bus and calculates the optical powers for the two lenses, using the distance data and the patient-specific correction algorithm. These two optical powers are then mapped to the corresponding voltage values, using the optical power- actuation voltage relation of the variable-focus lens, shown in Fig. 7 (b). The master microcontroller sends these two voltage values to the slave microcontroller through a standard UART bus, operating at 19,200 baud. The slave microcontroller is responsible for continuously converting the voltage data to pulse-width-modulated (PWM) control signals, which digitally control the analog output of the ultra-miniature DC-DC piezo transformer. A resistor-divider feedback network provides real-time measurements of the lens voltage, allowing our system to be precisely and power-efficiently controlled.

Prescriptions of patients are bound to change as time and presbyopia progress within their ocular systems. It has also been shown that visual acuity and accommodation response are dependent on illumination [102], [116]–[118]. We incorporated a Bluetooth Low Energy (BLE) module within our system, which works in conjunction with a smartphone application. This allows the patients to easily enter their updated prescriptions and user settings into the eyeglasses' control electronics. This BLE subsystem is connected to the master microcontroller through its principal hardware SPI bus, enabling over-the-air (OTA) upgrades for the entire system-controlling software, when required.

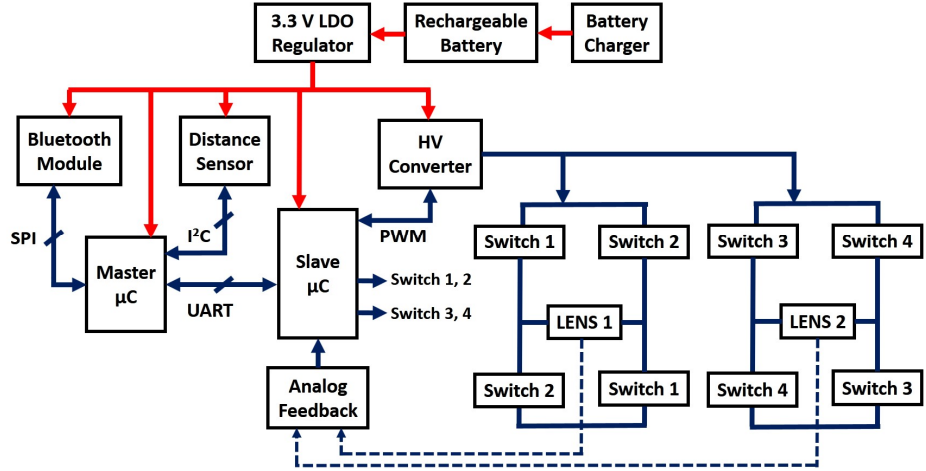

Fig. 8. Block diagram of the entire autofocusing system.

### B. Power Consumption

In practice, bulky power packs or batteries to reliably power sophisticated presbyopia correction systems for long hours, are unfeasible, inconvenient, and unacceptable to users. Presbyopes require *continuous* accommodation correction and such technologies must stay operational, preferably without charging, to allow the users unimpeded functionality. We specifically utilize low power microcontrollers (Microchip ATmega32U4) and BLE SoC (Nordic nRF51822) in our design to facilitate better power efficiency. Key systems in the master microcontroller which are not used, like the analog-digital converter (ADC) and a few timers have been disabled to conserve power. The microcontrollers run with a clock frequency of 8 MHz to save power, without affecting the computation speed. The entire system is operated with a 3.7 V, 400 mAh, rechargeable lithium-ion polymer battery. We have also incorporated a battery charger (Microchip MCP73831) and a fuel gauge (Texas Instruments bq27441) which measures the remaining battery capacity and communicates battery status with the master microcontroller through a high speed $I^2C$ bus. Power is routed to the entire electronics system through a 3.3 V, low-noise, low dropout voltage regulator (MaxLinear SPX3819). A low power on/off switch (Maxim Integrated MAX16054) in conjunction with a push button constitute the power switching subsystem. We measured the current consumption of our system at 3.3 V with a 6 ½ digit multimeter (Agilent 34401A), to be ~21 mA. This value increased to 22 mA, during BLE communications with a smartphone application. However, software and prescription upgrades, which require BLE functionality, are rare and hence, the average steady-state current consumption of 21 mA can be safely approximated for this system. With the 400 mAh capacity lithium-ion polymer battery we use in our design, this ideally yields an average operational time of ~19 hours between every charge cycle, when the system is configured to perform 1 correction per second. However, battery age and frequency of use will determine the practical limits of this operational time.

### C. Accommodation Correction and Image Quality

The autofocusing system's ability to provide adequate optical correction in presbyopes was evaluated by comparing their AD before and after optical correction with our system. Fig. 9 (a) shows the deficiency models (solid lines) and deficiency data (markers) for 5 different eyes of uncorrected presbyopes, presented in [102]. The increase in their AD corresponding to increasing stimulus (i.e., decreasing object distance) can be clearly seen from the deficiency curves in Fig. 9 (a). The AD increases from ~0 D to >3.0 D as the stimulus is progressively increased from 1.0 D (object distance = 1.0 m) to 3.3 D (object distance = 30 cm). Ideally, optical corrections using autofocusing eyeglasses should reduce this stimulus-dependent deficiency in presbyopes to 0 D. Fig. 9 (b) shows the deficiency in the same 5 presbyopia patients after correction using our autofocusing system. The curves represent difference in the optical correction determined by the patient-specific AD models and our system's response measured at 5 object distances (1 m, 70 cm, 50 cm, 40 cm and 30 cm). Our results indicate that the average corrected AD for the 5 patients ranged from -0.021 D to 0.016 D over the 5 object distances, using our autofocusing system. The maximum corrected deficiency in presbyopes did not exceed 0.4 D. The system was able to repeatedly generate the optical powers as determined by the presbyopia correction algorithms for different patients.

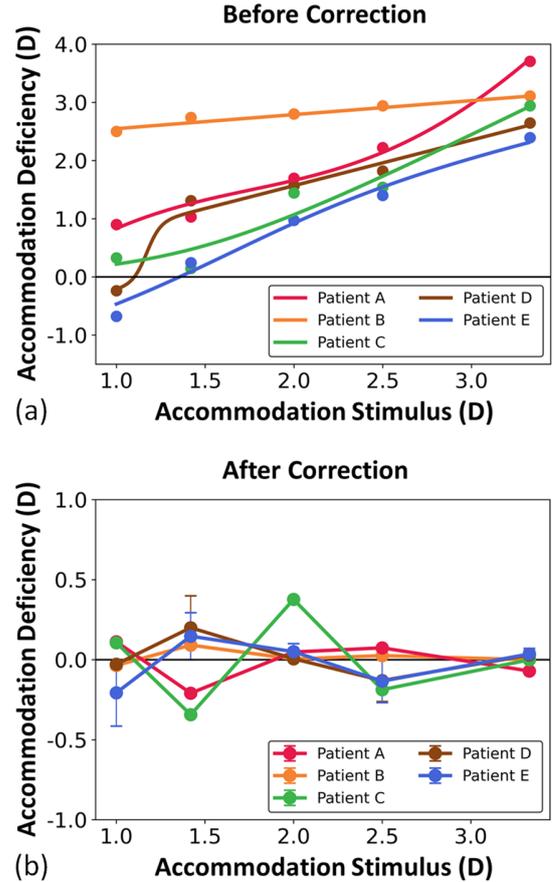

Fig. 9. (a) Accommodation deficiencies of 5 patients before correction, plotted using models developed in [102]. (b) Average corrected accommodation deficiencies of the same 5 patients after our autofocusing system correction. Bars represent the standard deviation. Corrected AD does not exceed 0.2 D.

Optical resolution plays an important part in determining the effectiveness of autofocusing systems to be used for presbyopia correction. To that extent, we also measured the approximate, subjective optical resolution of this system with the help of a 1951 USAF optical resolution test chart. Fig. 10 (a)-(c) show the images of the resolution test chart placed at different distances away from variable-focus lens. The autofocusing system was running a patient-specific algorithm, developed in [102], and automatically set the appropriate optical powers (necessary to correct that patient's AD) on the variable-focus lens. We observed the smallest, clearly visible elements on the test chart. Our observations show that the resolution of this system was between 0.75 – 1.00 lp/mm.

An autofocusing system intended to provide optical correction to a presbyope should also have optical quality comparable to that of an average human eye. We measured the 1-D vertical and horizontal MTFs for the variable-focus lens and compared it to that of a best-corrected human eye. An analytical formula constructed by Watson [119] was used to model the pupil-diameter dependent MTFs for a human eye. Fig. 11 compares the 1-D MTFs, measured for our lens to those modelled for a best-corrected human eye. We report an average MTF-50 (i.e., spatial frequency at 50% MTF) value of 10.5 cycles/degree for our variable-focus lens.

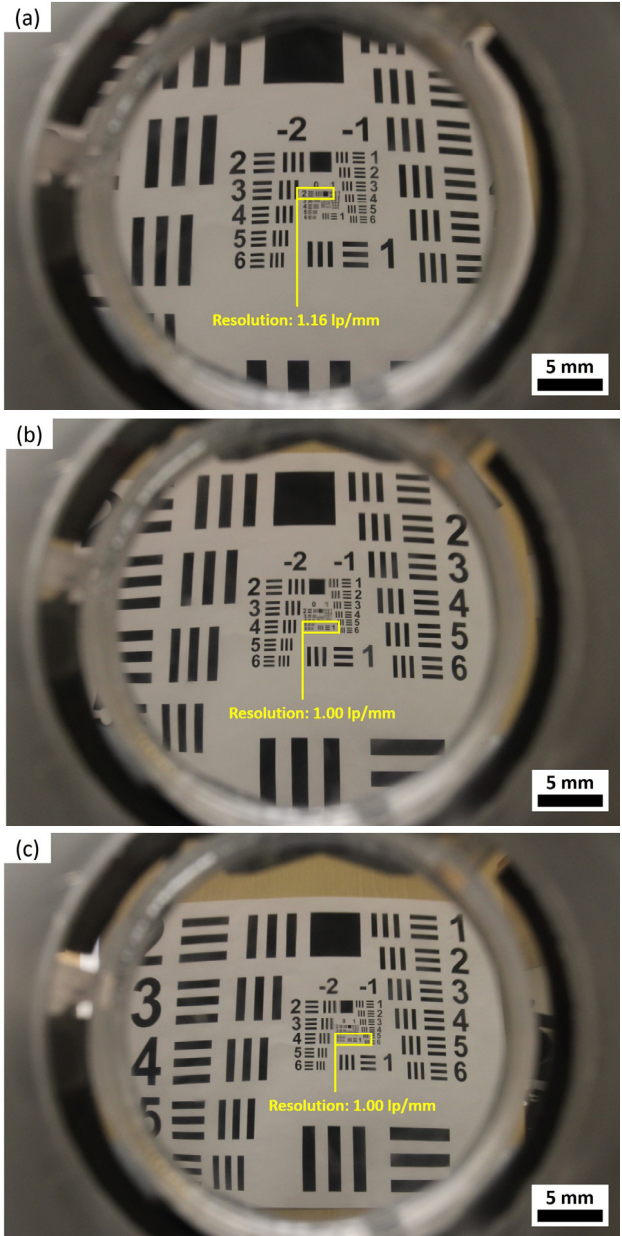

Fig. 10. Photos of the USAF optical resolution test chart through the variable-focus lens, placed at (a) 30 cm, (b) 40 cm and (c) 50 cm. The accommodation deficiency model for patient no. 3 was used in conjunction with the autofocusing system for these images [102]. The photos highlight the smallest, clearly visible element on the optical resolution test chart.

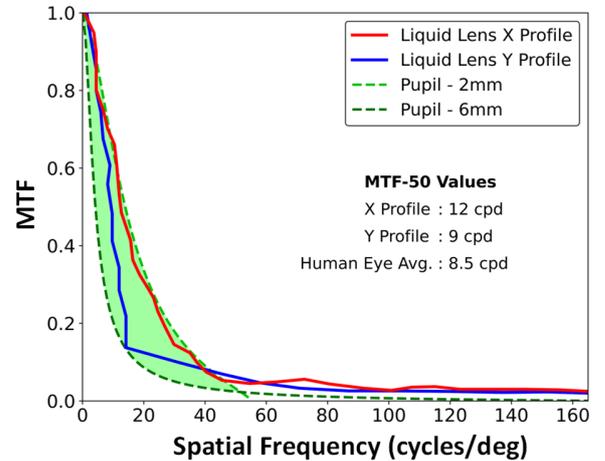

Fig. 11. Comparison of the 1-D MTF of variable-focus lens used in the autofocusing system and the radial MTF of the human eye. The solid red and blue lines represent the 1-D MTFs of the variable focus lens along the X (horizontal) and Y (vertical) axes, respectively.

The green shaded area represents the range of the MTFs for a best corrected human eye.

*D. Response Time*

The electronics control system and the lenses introduce inherent time delays in the autofocusing system. As reported before, the lenses have a response time of ~40 ms, i.e., it takes

~40 ms from the moment a particular voltage is applied to the actuators of the lenses till the point where the lenses produce the expected optical power.

Our measurements show that the average time taken by the electronics subsystem to produce the required voltages for the lenses is ~67 ms. This response time takes into account the time required by the distance sensor to complete its ranging operation, the time required by the master microcontroller to extrapolate the correct optical powers and their related voltages, and the time required by the slave microcontroller to set these voltages across the lens actuators. Hence, the total time taken by this system from the moment distance ranging begins, till the point where the lenses produce the appropriate optical power, is 107 ms. This enables us to have a maximum refresh rate of ~10 Hz, i.e., the system can change the optical powers of the lenses up to 10 times a second.

### E. Ergonomics, Aesthetics and Portability

Autofocusing systems are meant to be worn all the time, throughout the daily routine of a presbyope, in order to provide seamless accommodation. In practical terms, this requires the system to be lightweight, comfortable and fashionable enough to be worn during daily engagements. Engineering limitations impose strict restrictions on the sizes of various components within our system, for example, the lenses cannot be thinned down any further, without sacrificing their electro optical performance. Similarly, the housing for electronics cannot be any thinner without sacrificing battery capacity. Despite such limitations, our system attempts to provide a frame design which is much *sleeker* compared to recent similar technologies [98], [100], [101]. Fig 12 (a)-(d) show the mechanical dimensions of the autofocusing eyeglasses.

Fig. 13 (a) shows the arrangement of the internal components in these eyeglasses. The maximum *thickness* of the frame is 14.46 mm, while the maximum *height* of the electronics housing is 28.00 mm. The diameter of the lens housing is 51 mm to house lenses with diameter 50 mm. The 0.5 mm annular gap between the lens and its housing helps in snugly fitting FPC communications cables and the HV wires for the lenses. The total weight of these autofocusing eyeglasses is 132.06 g. Our system has been designed to automatically focus on the object which is directly in the line-of-sight of the ToF distance sensor.

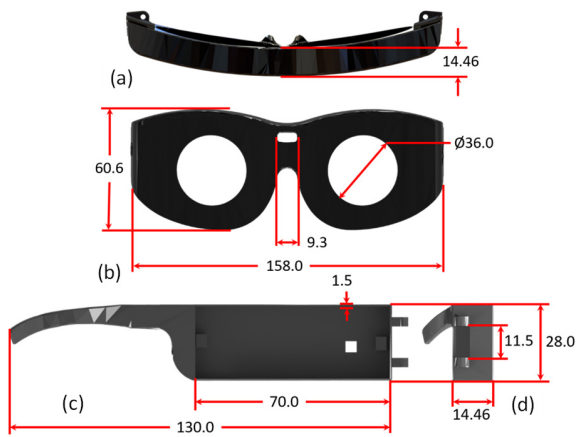

Fig. 12. Dimensions of the lens and electronics housing in the autofocusing eyeglasses system. (a) and (b) show the top and front views of the lens and ToF distance sensor housing, respectively, while (c) and (d) show the side and front views of the electronics housing, respectively. Dimensions are in mm.

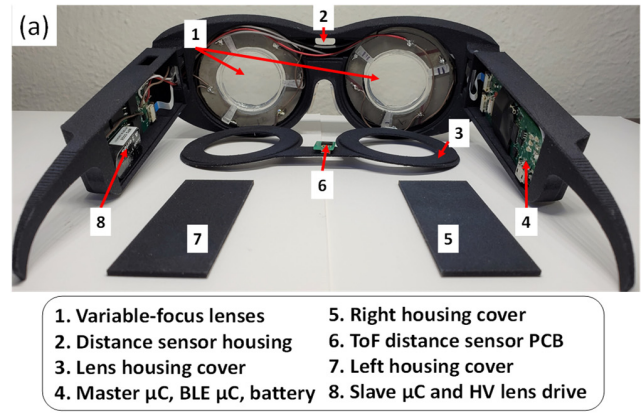

1. Variable-focus lenses
2. Distance sensor housing
3. Lens housing cover
4. Master μC, BLE μC, battery
5. Right housing cover
6. ToF distance sensor PCB
7. Left housing cover
8. Slave μC and HV lens drive

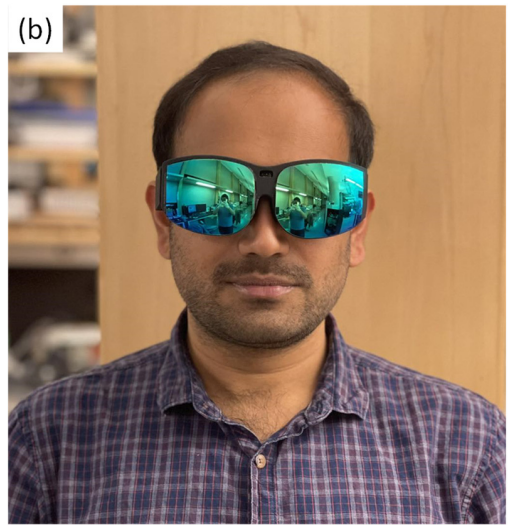

Fig. 13. (a) Autofocusing eyeglasses from the inside. This photo shows the various subsystems and their placements within the eyeglasses. (b) Ergonomic fit of autofocused eyeglasses resembles oversized sunglasses. The reflective covers hide the variable-focus lenses.

Some limitations exist for this autofocusing system. The eyeglasses do not change their optical power if the user changes their gaze direction, without moving their head. This can often lead to "jumps" in focus when the object distance changes drastically, in a short amount of time. However, such issues can be addressed with a low-profile, low-power digital oculometer designed for this system [120]. It should be noted that such upgrades will consume more power, leading to a faster battery discharge.

### IV. CONCLUSION

We demonstrated the design, assembly, and performance of an autofocusing eyeglasses system which can potentially restore accommodation in presbyopes. These autofocusing eyeglasses consist of two, variable-focus, liquid-filled lenses, a ToF distance sensor and a rechargeable battery-powered electronics control system, which utilizes patient-specific accommodation deficiency models to restore pre-presbyopic levels of accommodation in them. This system was evaluated on its ability to perform high fidelity and accurate optical corrections based on presbyopic patient data from a clinical trial. The optical resolution (MTF-50) of our autofocusing eyeglasses is 10.5 cycles/degree. Our system can provide up to 4.3 D in accommodation correction.